\documentclass[%
 aip,
 rsi,
 amsmath,amssymb,
 reprint,%
]{revtex4-1}

\usepackage{graphicx}
\usepackage{dcolumn}
\usepackage{bm}

\usepackage[utf8]{inputenc}
\usepackage[T1]{fontenc}
\usepackage{mathptmx}
\usepackage{etoolbox}
\usepackage{siunitx}

\makeatletter
\def\@email#1#2{%
 \endgroup
 \patchcmd{\titleblock@produce}
  {\frontmatter@RRAPformat}
  {\frontmatter@RRAPformat{\produce@RRAP{*#1\href{mailto:#2}{#2}}}\frontmatter@RRAPformat}
  {}{}
}%
\makeatother
\usepackage{xcolor}
\usepackage{wasysym}

\begin{document}

\preprint{AIP/123-QED}

\title{A high-speed, large-aperture 3D printed laser shutter for cold atom experiments}

\author{Samantha L. Rubinstein}
 \thanks{These authors have contributed equally to this work and share first authorship.}
\affiliation{School of Mathematics and Physics, University of Queensland, St. Lucia, QLD 4072, Australia.}

\author{Charles W. Woffinden}
 \thanks{These authors have contributed equally to this work and share first authorship.}
\affiliation{Australian Research Council Centre of Excellence for Engineered Quantum Systems, School of Mathematics and Physics, University of Queensland, St. Lucia, QLD 4072, Australia.}

\author{Maarten T. M. Christenhusz}
\thanks{These authors have contributed equally to this work and share first authorship.}
\affiliation{Australian Research Council Centre of Excellence for Engineered Quantum Systems, School of Mathematics and Physics, University of Queensland, St. Lucia, QLD 4072, Australia.}

\author{Guillaume Gauthier}
\email{g.gauthier@uq.edu.au}
\affiliation{Australian Research Council Centre of Excellence for Engineered Quantum Systems, School of Mathematics and Physics, University of Queensland, St. Lucia, QLD 4072, Australia.}
\author{Tyler W. Neely}
\affiliation{Australian Research Council Centre of Excellence for Engineered Quantum Systems, School of Mathematics and Physics, University of Queensland, St. Lucia, QLD 4072, Australia.}
\date{\today}

\begin{abstract}

Cold atom experiments require optical shutters with rapid and repeatable actuation. Existing solutions often involve high cost, limited aperture diameter and slow transitioning. We present a 3D-printable shutter design featuring a large aperture of 4.5~mm diameter, low-cost assembly, and robust performance. The device achieves switching times of $\SI{399\pm8}{\micro\second}$ with delays of {$\SI{2.7\pm0.2}{\milli\second}$} for a fully illuminated aperture, with four different shutters demonstrating repeatable performance. We measure the shutter speed to be $\SI{7.0\pm0.1}{\meter\per\second}$. This approach offers an accessible, scalable solution for cold atom and other optical experiments requiring multiple shutters with precise light control and avoids the need to resize or precision-align a beam through an aperture.

\end{abstract}

\maketitle

\section{Introduction}

Cold atom experiments require precisely controlled sequences of light pulses for the cooling and manipulation of atoms. These sequences are generally created using transistor-transistor logic (TTL) to operate acousto-optical deflectors (AODs) for fast transition, and shutters for full extinction. The required periods of pulses can range from the order of milliseconds needed to image atoms to seconds over which a magneto-optical trap (MOT) is formed. In order to perform useful science, the sequence must be highly repeatable, so that the same atomic conditions are obtained with each experimental run. This requires shutters that have fast and repeatable operation whilst maintaining a small footprint to fit within crowded optical setups. 

There are various approaches to designing shutters which have been integrated into such experiments. Off-the-shelf solutions are available, but are typically expensive in the quantities required by Bose-Einstein condensation experiments (typically 10 or more). Our own experiment (described in detail in Ref. \onlinecite{Gauthier_2016}) has utilized the ``iShutter" system of Maguire et al. which uses the voice actuators from iPod hard drives as the shutter mechanism.\cite{Mitchell_2008, Maguire_2004} These have been found to be very reliable, but are based on now obsolete hardware. Recently, Zhang et al. demonstrated the use of 3D printed shutters driven by a DC motor.\cite{Zhang_2015} This design features a small-diameter (2.5 mm) aperture, which produces short delay and transition times. However, with typical $\mathrm{1/e^2}$ laser beam diameters of 1--3 mm,\cite{Top_TA, Mog_OA,Spectra_Mi} precision alignment of the shutter is required and can potentially cause clipping of the beam without focusing it through the shutter, \cite{Scholten_2007,Singer_2002} or reducing the beam diameter with a telescope, adding to the cost, complexity and size of an optical setup.

We attempted to recreate the original design of Zhang et al. with a larger diameter aperture, in order to avoid these issues, but we were unable to match the performance parameters achieved with our current system. For this reason, we refined the shutter design, maintaining the convenience and accessibility of the almost entirely 3D printed structure and small DC motor, but adapting the design to accommodate a larger aperture diameter (4.5 mm) while matching or exceeding the performance obtained with our current iShutter system. We also assess the variability in assembly, a potential issue posed by the nature of 3D printing and manual assembly. 
Consistency is important as multiple shutters are often required for a single experiment, and researchers need to reliably predict (and hence account for) the lag in beam switching introduced by shutters. It can become impractical to do this when performance varies greatly between shutters.  

The three main metrics used to characterize a shutter are illustrated in Fig.~\ref{fig:parameters} and are described herewith:

\begin{figure}
    \centering
    \includegraphics[width=1.0\columnwidth]{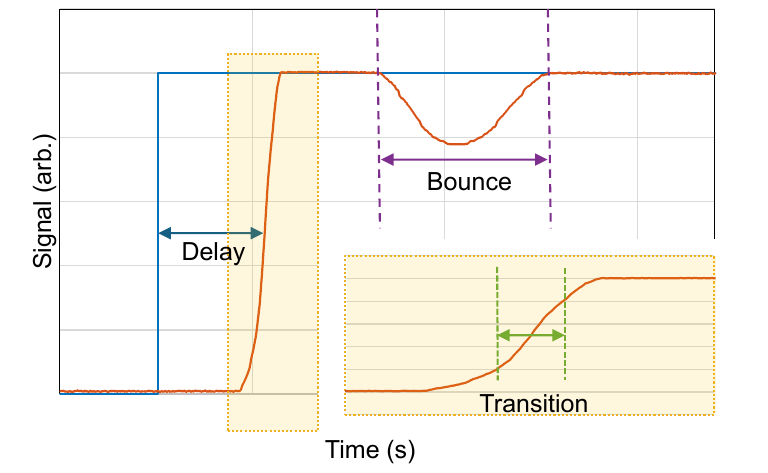}
    \caption{Illustrative graph showing delay time and bounce parameters. The blue line shows the TTL signal and the orange line shows the photodiode response to illumination after the shutter. Insert: reduced time-base view of photodiode response of shutter to illustrate the transition time.}
    \label{fig:parameters}
\end{figure}

\begin{enumerate}
\renewcommand{\labelenumi}{(\roman{enumi})}
  \item  Delay time: The time between the signal being sent to open/close the shutter and the shutter opening/closing. This time should be reasonably short, but most importantly, must be consistent. A consistent delay ensures that the experimental sequence remains repeatable. As long as the delay is reproducible and does not exceed the pauses between shutter transitions in the experimental sequence, it can be compensated for in software. Delay times reported herewith are calculated to the point at which the signal level reaches 50\% of the open/closed value. 
  
  \item Transition time: The time between the shutter beginning to open/close, and being completely open/closed. This time must be as short as possible to ensure precise control of experiments, and must also be repeatable. It is calculated as the time it takes from the shutter signal to transition from 20-80\% or \textit{vice versa}.
  
  \item Bounce: The amount of back-movement the shutter suffers after hitting the stop. Any bounce which causes the shutter to temporarily block/unblock the aperture must be eliminated. 

\end{enumerate}

Here, we describe the design of our shutter and detail its performance in relation to the metrics listed above. We compare performance between four shutters of the same design to determine the repeatability of the design and manufacturing process.

\section{Design and Assembly}
Scaling up the design presented by Zhang et al.~\cite{Zhang_2015} for a larger diameter aperture, whilst maintaining the compact form factor, introduces challenges such as increased shutter blade mass and consequently longer transition and delay times. Our design addresses these limitations by optimizing both the shutter mechanism and drive electronics for a larger diameter aperture, guided by the performance metrics outlined in the introduction.

\subsection{System Overview}
\begin{figure}
    \centering
    \includegraphics[width=0.9\columnwidth]{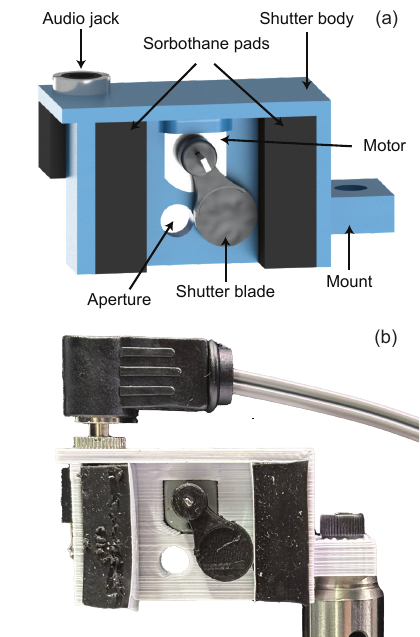}
    \caption{(a) Annotated CAD render of the shutter. (b) Photo of the shutter mounted on an optical post. Electrical connections between the motor and audio jack exist at the rear of the shutter. Schematics and CAD files are provided within Ref.~\onlinecite{rubinstein_2026}.}
    \label{fig:CAD}
\end{figure}

The shutter operates based on a TTL control signal: a high level (3.3~V) closes the shutter, while a low level ($\sim 0$~V) opens it. The TTL signal drives an opto-coupler, which controls an H-bridge to reverse current through a DC motor connected to an RC network (see Sec.~\ref{sec:electronics}). The motor rotates a shutter blade that either blocks or clears the aperture within the shutter body. Mechanical stops on either side of the aperture define the open and closed positions (Fig.~\ref{fig:CAD}).

\subsection{Mechanical Design}

We 3D printed the shutter body using PLA+ for reasons of cost and convenience. The shutter blade was printed in Onyx\copyright, a composite material with a high heat deflection temperature of $145~^\circ$C, which helps prevent deformation when exposed to higher-power laser beams. It had a thickness of 0.8~mm at its contact point and weighs 0.15~g.

Transition time across the aperture is strongly influenced by aperture diameter: smaller apertures allow shorter travel distances and thus faster transitions. Enlarging the aperture diameter inherently increases transition time, necessitating improvements in both mechanical and electronic subsystems to maintain high speed. To maximize acceleration, the blade was designed to be as light as possible, with minimal thickness in the region shielding the aperture.

Delay time depends on the distance between the stop position of the blade edge relative to the aperture in the fully open and closed positions. Reducing this distance minimizes delay but can increase bounce. Similarly, increasing the blade speed reduces delay but also increases the back-action from the blade stop. These competing factors required careful optimization.

To prevent bounce, the material that the blade contacts must absorb as much kinetic energy as possible to prevent an impulse in the opposite direction. This is achieved by using a damping material. We selected a soft, 30-duro Sorbothane pad for this purpose---a viscoelastic polymer---as it is highly effective at damping vibration. Self-adhesive tape is used to cover the Sorbothane surfaces that comes into contact with the blade to prevent it sticking. Additionally, having a thinner blade surface making contact with the Sorbothane also reduces the bounce. These measures allowed for much improved transition and delay times without introducing bounce.

The blade diameter is 8.5~mm, significantly larger than the 4.5~mm aperture diameter. This oversizing slightly increases delay time, as the stops need to be located further from the aperture, but ensures that any residual motion after impact does not expose the aperture, preventing optical leakage.

A mounting point is included which allows the shutter to be fixed to a standard 12~mm diameter optical post.
The shutter system is both fast to assemble and low-cost. Details of the individual components and assembly instructions are provided as part of the dataset (Ref.~\onlinecite{rubinstein_2026}).

\subsection{Electronics Design}\label{sec:electronics}
The drive electronics design has been adapted from that of Maguire et al.\cite{Maguire_2004}, Zhang et al.\cite{Zhang_2015}, and our current shutter driver. The RC network provides a large initial current from a charged capacitor to overcome the motor’s back electromotive force (emf). 
During initial acceleration, the rate of change of the current through the motor, $dI/dt$,  is set by the voltage, which determines the back emf on the motor and thus the the initial acceleration. The high capacitance allows for an extended application of the increased torque to the motor. Once the capacitor is sufficiently discharged, the resistor supplies a small, constant current to hold the blade in position. 

\begin{figure}
    \centering
    \includegraphics[width=1\columnwidth]{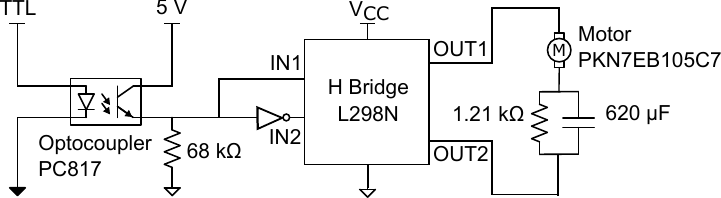}
    \caption{Electronic schematic of the shutter drive circuit. Further component details are provided in Ref.~\onlinecite{rubinstein_2026}.}
    \label{fig:circuit}
\end{figure}

For the duration of the shutter transition the capacitor delivers the bulk of the current to the motor through initial discharge and later charge, which is still ongoing when the blade impacts the Sorbothane pad. This counteracts the bounce and suppresses it. The behaviour is reflected directly in the elimination of bounce reported in Sec.~\ref{sec:stability}. The resistor provides a residual current to the motor to maintain the blade's position. A large resistance lowers the power consumption when the shutter is stationary. An optocoupler was incorporated into the drive electronics to protect TTL control signals from potential transients during switching. The circuit schematic is presented in Fig.~\ref{fig:circuit}.

For practical integration, the motor is connected via a 3.5~mm audio jack mounted on the shutter body, allowing easy connection and disconnection. Two-core, 24~AWG audio cable was used, rather than BNC coaxial cable due to its smaller profile and greater flexibility, allowing it to be easily routed through an optical bench setup. We did not observe any issues with cross-talk between shutter lines during testing. 

\begin{figure}
    \centering
    \includegraphics[width=1\columnwidth]{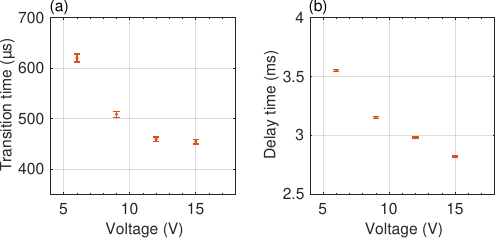}
    \caption{Shutter performance for fall transition times at different applied H-bridge supply voltages (${V_\mathrm{CC}}$). (a) is transition time, (b) is delay time. Standard errors calculated from repeat measurements are displayed.}
    \label{fig:voltages}
\end{figure}

\begin{figure}
    \centering
    \includegraphics[width=1\columnwidth]{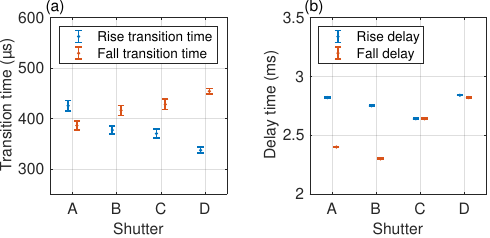}
    \caption{Performance comparison between different shutters operated at 15~V, (a) shows rise and fall transition times, while (b) shows rise and fall delay times. Data are laterally offset for clarity.}
    \label{fig:shutter_comparison}
\end{figure}

\section{Performance}
\subsection{Setup}
For all timing measurements, the shutter aperture was illuminated with an array of white LEDs, slightly focused by a lens to remove non-uniformity in intensity at the aperture. The light passes through a square mask measuring 10~mm~$\times$~10~mm prior to the shutter aperture to prevent leakage around the shutter. This configuration provided uniform illumination across the shutter aperture. Illuminating the entire aperture represents a worst-case operational condition as any residual bounce or incomplete extinction would be observable at any point across the aperture. After passing through the shutter aperture, the light was focused onto a fast photodiode. The photodiode output was compared to the TTL control signal using an oscilloscope allowing direct measurement of delay time, transition time, and any bounce. 

\subsection{Operating Voltage Optimization}
The shutter was driven using a TTL control signal operating at \SI{0.5}{\hertz} with a 50\% duty cycle. The H-bridge supply voltage ($V_{\mathrm{CC}}$) was varied in \SI{3}{\volt} increments from 9--15~V using a laboratory power supply. Note that prior testing with this design demonstrated that repeated operation at 18~V resulted in damage to the motor. For each voltage setting, rise and fall transition times and delay times were recorded for 1,500 consecutive cycles using a single shutter (Shutter D). Results are presented in Fig.~\ref{fig:voltages}. 

It is apparent from Fig.~\ref{fig:voltages} that performance improves with increasing voltage, but the improvement diminishes with increasing voltage. For this reason, we selected 15~V as the normal operating voltage.

\subsection{Reproducibility and long term stability}\label{sec:stability}

To assess reproducibility across four different shutters, all performance parameters were measured at 15~V for each unit over 1,500 consecutive cycles. The results are summarized in Fig.~\ref{fig:shutter_comparison}, which demonstrates close agreement between shutters for both transition and delay times, indicating that the design and assembly process yields consistent performance.

\begin{figure}
    \centering
    \includegraphics[width=1.0\linewidth]{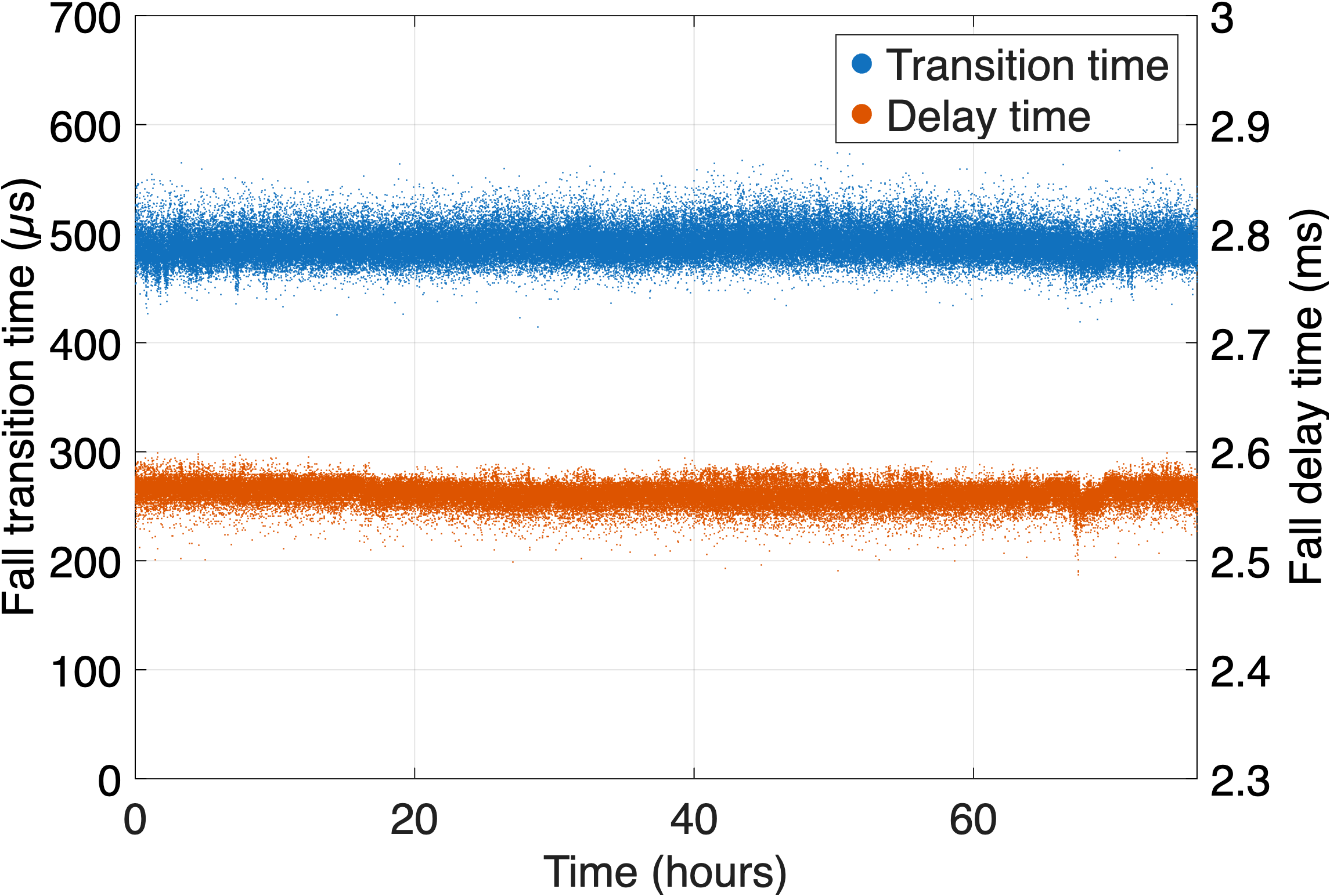}
    \caption{Long-term performance for one shutter over 120,000 close transitions. Performance is consistent across the time period.}
    \label{fig:Long_Time}
\end{figure}

Long-term stability was evaluated by operating a single shutter continuously for approximately 120,000 close transitions at 15~V. Figure~\ref{fig:Long_Time} shows the measured delay and transition times over this interval. No systematic drift or degradation in performance was observed, confirming the suitability of the shutter for extended use.

We compare the results from our shutter to those acquired with other research groups' designs in Table~\ref{tab:results}.
\begin{table*}
\caption{\label{tab:results}Comparison of performance metrics between the shutter presented in this work to other shutter designs. Transition time is strongly dependent on beam size. Transition times are 20-80\% unless stated. Standard deviations in each case are for a single shutter.}
\begin{ruledtabular}
\begin{tabular}{ccccccc}
 Shutter Design & Spot size & Delay & Stdev & Transition & Stdev & Comments  \\ \hline
Zhang\cite{Zhang_2015}       & 1~mm & 2.6~ms & \SI{2}{\micro\second} (0.08\%) & \SI{244}{\micro\second} & \SI{2}{\micro\second} (1\%) & - \\ 
Fischer\cite{Fischer_2024}    & 2~mm & <50~ms & - & 3.2~ms & 0.3~ms & opening times \\ 
Maguire\cite{Maguire_2004}   & \SI{30\pm1}{\micro\meter}  & 6.06~ms & \SI{87.4}{\micro\second} & \SI{6.2}{\micro\second} & \SI{0.3}{\micro\second} & 10-90\% transition, 16V operation \\ 
Mitchell\cite{Mitchell_2008} & \SI{300}{\micro\meter}  & 3.31~ms & \SI{3.66}{\micro\second} & \SI{410}{\micro\second} & \SI{3.23}{\micro\second} & 10-90\% transition \\ 
Reported here                & 4.5~mm & \SI{2.65}{\milli\second} & \SI{0.2}{\milli\second} & \SI{399}{\micro\second} & \SI{8}{\micro\second} & Uniform illumination  \\ 
Reported here                & 1~mm & &  & \SI{60}{\micro\second} &\SI{1}{\micro\second} &  Calculated for 1 mm Gaussian beam (Supplementary Material, Sec.~\ref{sec:sup})\\ 
\end{tabular}
\end{ruledtabular}
\end{table*}

Our shutter notably has the largest illumination spot size and one of the shortest delay times. Our design exhibits extremely fast transition times under full aperture illumination, comparable to results from shutters with much smaller illumination sizes. When our shutter's performance is normalized for a 1~mm Gaussian beam, (see Supplementary, Sec.~\ref{sec:sup}) the equivalent transition time is $\SI{60\pm1}{\micro\second}$ --- a factor of 4 improvement compared to that of Zhang et al. and is only surpassed by that of Maguire et al. with a \SI{30\pm1}{\micro\meter} spot size. This is achieved without the onset of bounce, whilst reducing the risk of obsolescence by using generic, non-specialized parts.

\subsection{Shutter Speed Measurement}
The shutter blade velocity was measured by placing an additional rectangular mask of dimensions 10~mm $\times$ 1.0~mm between the aperture and the blade. As the blade sweeps across the mask, the transmitted intensity varies approximately linearly with blade displacement. The photodiode signal thus provides a direct measure of the instantaneous blade position as a function of time.

Using this approach, we calculated the shutter speed for a single shutter as $\SI{7.0\pm0.1}{\meter\per\second}$. This compares to \SI{1.22\pm0.02}{\meter\per\second} from Ref.~\onlinecite{Zhang_2015} and \SI{10.1\pm0.1}{\meter\per\second} from Ref.~\onlinecite{Maguire_2004}, when using a much higher supply voltage of 30~V.

\subsection{Electrical model}
We created a SPICE model of the circuit to determine its electrical properties,\cite{rubinstein_2026} and have used it to calculate the instantaneous current through the motor at 15~V as $\sim$5.9~A, with an average power dissipation per motor of $\sim$560~mW when operated at 1~Hz. The current required to maintain the blade in either the open or closed position whilst also preventing bounce is $~\sim12$~mA (180~mW).
 The model indicates that upon reverse bias from the H-bridge, the capacitor will charge to 99\% of its final voltage within $\sim$18~ms corresponding to 96\% of its initially stored energy. This time represents a conservative minimum for repeatable switching operation.

\subsection{Thermal Performance}
To determine the robustness of the shutter to typical illumination conditions, we exposed the shutter blade to a 780~nm laser beam with a 1.3~mm beam-waist for a range of powers. After 48 h of continuous exposure at a power of 125~mW, no visible deformation or softening of the blade was observed. At a slightly higher power of 150~mW with 2 minutes of exposure, the surface finish of the blade changed, indicating that the heat deflection temperature had been reached. This demonstrates that the shutter can withstand $>$100~mW optical power from a beam waist representative of typical cold-atom laser optical setups.

For applications where higher powers are required, the principal limitation of the Onyx shutter is that the low thermal conductivity of the blade material causes heat to remain highly localized at the laser beam spot. The thermal limit could be improved by using Al foil, fixed to the surface of the blade. This would more effectively conduct heat across the blade without adding significantly to the weight. To reduce the risk of spurious specular reflections from the beam, the foil could be sand-blasted to create a more diffuse reflection profile. 

A simple steady-state thermal model of the blade shows that a power of ~235~mW uniformly distributed across the full circular area of the blade is sufficient to raise the blade temperature to the heat deflection temperature of Onyx. Improving thermal conduction, for example with Al foil, could therefore raise the power limit from ~125 mW towards ~235 mW, beyond which a different blade material would be required. Our thermal model can be found in Ref.~\onlinecite{rubinstein_2026}.

\section{Conclusions}
We have demonstrated and characterized a low-cost, 3D-printable laser shutter optimized for cold atom experiments that combines a large aperture diameter with fast and repeatable actuation. The shutter provides a clear aperture of 4.5~mm ---much larger than other designs --- while achieving transition times of $\SI{399\pm8}{\micro\second}$ and delay times of $\SI{2.7\pm0.2}{\milli\second}$ under worst-case full-aperture illumination. When operating with a 1~mm Gaussian laser profile, we achieve an equivalent transition time of $\SI{60\pm1}{\micro\second}$. This design significantly outperforms existing smaller aperture designs, without requiring minification of the beam -- requiring additional optics and precision alignment. Performance was found to be highly repeatable across four independently assembled shutters and was stable over more than 120{,}000 switching cycles, which confirms the suitability of the materials used in such a design.

By introducing Sorbothane within the mechanical design as a shock absorber, and optimizing the RC circuit drive electronics, shutter bounce was eliminated without sacrificing speed. The shutter blade velocity was measured as $\SI{7.0\pm0.1}{\meter\per\second}$. Compared to previously reported mechanical shutters, the present design achieves competitive transition and delay times while operating with significantly larger illumination spot sizes, thereby avoiding the need for beam resizing or precision alignment.

The compact form factor, low power consumption, and a combination of off-the-shelf and 3D-printed components make this shutter well suited for experiments requiring multiple independently controlled shutters in compact optical arrangements. We have provided design files, circuit schematics and a parts list in Ref.~\onlinecite{rubinstein_2026} to allow adoption and further modification by other laboratories.

Future modifications could extend the design to higher optical powers by using different materials and surface finishes of the blade to either enhance thermal dissipation, or survive higher temperatures. 

\section*{Acknowledgements}
S.L.R. acknowledges the support of the University of Queensland, School of Mathematics and Physics Winter Research Program in providing a scholarship for this activity. C.W.W. acknowledges support from an Australian Army Exploit Project grant. M.T.M.C acknowledges the support of Australian Research Council Discovery Project No. DP250103551. G.G. acknowledges the support of the Queensland Government through the Quantum 2032 Challenge Program. \\

\section*{Author Declarations}
\subsection*{Conflict of Interest Statement}
The authors have no conflicts to disclose.\\

\subsection*{Author Contributions}
\noindent \textbf{S. L. Rubinstein:} Data curation (supporting), Formal analysis (lead), Investigation (lead), Writing – original draft (supporting), review and editing (equal) \textbf{C. W. Woffinden:} Conceptualization (supporting), Data curation (lead), Formal analysis (supporting), Methodology (lead), Investigation (supporting), Software (lead), Writing – original draft (lead), review and editing (equal)
\textbf{M. T. M. Christenhusz:} Data curation (supporting), Investigation (supporting), review and editing (equal), software (supporting)
\textbf{G. Gauthier:} Conceptualization (lead), Methodology (supporting), review and editing (equal), investigation (supporting)
\textbf{T. W. Neely:} writing – review and editing (equal), Supervision (lead)
\section*{Data Availability}
The data and models, including experimental data, CAD files, a parts list, assembly instructions and thermal and electrical models that support the findings of this study are openly available in UQ eSpace at https://doi.org/10.48610/da433e5 (Ref.~\onlinecite{rubinstein_2026}).

\section*{Supplementary Material} \label{sec:sup}
\subsection{Calculation of transition time for a 1~mm Gaussian beam}
For a Gaussian beam, transmission as a function of a straight-edged shutter blade is given by
\begin{equation}
    T(x)=\frac{1}{2}\mathrm{erfc}\frac{\sqrt{2}x}{w}
\end{equation}
where erfc is the complementary error function and $w$ is the beam waist.

At 80\% transmission
\begin{equation}
0.8=\frac{1}{2}\mathrm{erfc}\left(\frac{{\sqrt{2} x_{80}}}{w}\right).
\end{equation}
To calculate the value of $x_{80}$, we use the inverse complementary error function (erfcinv):
\begin{equation}
\mathrm{erfcinv}(1.6)=\frac{\sqrt{2}x_{80}}{w}.
\end{equation}
For a beam diameter ($\mathrm{1/e^2}$) of 1 mm (beam waist, $w=\SI{0.5}{\milli\meter}$, this gives $x_{80}=-0.21$~mm.\\
Due to symmetry, $x_{20}=-x_{80}$, so the distance the blade travels for 80\% to 20\% transmission is $0.84w = \SI{0.42}{\milli\meter}$.
Using the previously measured blade speed of \SI{7.0}{\meter\per\second}. The transmission time is thus 
\begin{equation}
    \frac{0.42}{7.0}~\mathrm{ms}=60~\mathrm{\mu s} 
\end{equation}

\section*{References}
\bibliography{refs}

\end{document}